\documentclass[preprint,showpacs,aps]{revtex4}
\usepackage{graphicx}
\usepackage{dcolumn}

\usepackage[english]{babel}

\begin{document}

\title{Regge analysis of diffractive and leading baryon structure functions 
from DIS}
\author{M. Batista}
\author{R. J. M. Covolan}
\author{J. Montanha}\altaffiliation{Work supported by Fapesp, Proc. 99/01236-9}
\affiliation{Instituto de F\'{\i}sica {\textit Gleb Wataghin} \\
Universidade Estadual de Campinas, Unicamp \\
13083-970\ Campinas \ SP \ Brazil}

\date{\today}

\begin{abstract}
{In this paper we present a combined analysis of the H1 data on
leading baryon and diffractive
structure functions from DIS, which are handled as two components of
the same semi-inclusive process. The available
structure function data are analyzed in a series of fits in which three main
exchanges are taking into account: pomeron, reggeon and pion. For each of
these contributions, Regge factorization of the correspondent structure
function is assumed. 
By this procedure, we extract information about the interface
between the diffractive, pomeron-dominated, region and the leading
proton spectrum, which is mostly ruled by secondary exchanges.
One of the main results is that the relative reggeon contribution to
the semi-inclusive structure function is much smaller than the one
obtained from a analysis of the diffractive structure function alone. }
\end{abstract}


\maketitle

\section{\label{Intro}Introduction}

One of the most striking results obtained at the DESY HERA $e p$ collider was 
the discovery by the H1 and ZEUS collaborations \cite{H1-old,ZEUS-old} that  
deep inelastic scattering (DIS) events tagged with rapidity gaps exhibit mass  
distributions whose shape resemble very much those observed
in hadron-hadron diffraction experiments.
More recently, both the H1 and ZEUS collaborations reported 
\cite{H1-Lead,ZEUS-Lead} analyses 
of another class of DIS events whose pretty flat distribution turned
out to be  
quite similar to the {\textit leading particle} spectrum, also 
observed in hadron  
reactions. These similarities suggest that the Regge pole 
phenomenology \cite{Collins}, successfully used to  
describe diffractive events and the leading particle effect in hadron  
processes \cite{Collins,Mirian}, might also be employed to analyze the  
corresponding events obtained in DIS.

\begin{figure}[ht]
\includegraphics[height=16cm]{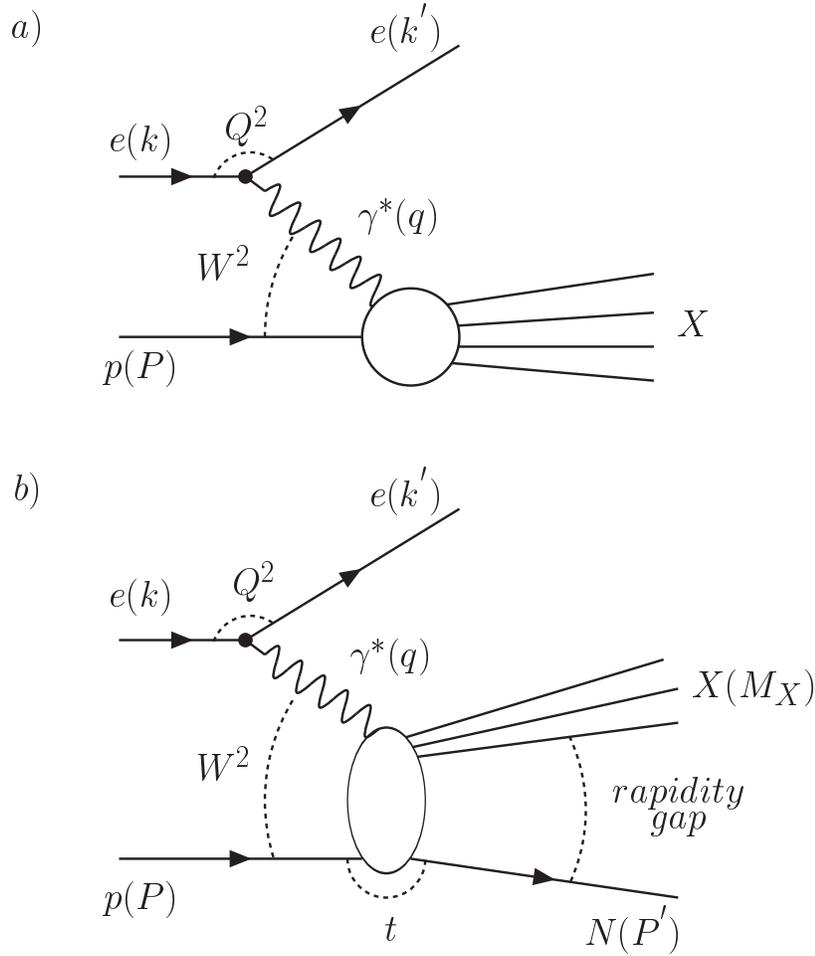}
\caption{\label{fig1}a) Kinematic variables for the reaction 
$e\;p \rightarrow e\;X$;
b) Kinematic variables for the semi-inclusive reaction $e\;p
\rightarrow e\;N\;X$, where $N$ stands for a proton or neutron.}
\end{figure}
  
In a conventional DIS process, $e\,p\rightarrow e\,X$, a high energy electron  
of four-momentum $k$ interacts 
with a proton of four-momentum $P$ through the emission of a photon of 
virtuality $Q^{2}$. As long as the photon has high enough momentum, it can 
resolve the internal partonic structure of the proton, interacting with its  
partons through a hard scattering which breaks up the hadron. 
In this inclusive reaction only the outgoing electron is detected in the 
final state (Fig.~\ref{fig1}a). 
 
If, besides the electron, one specific kind of hadron is detected in
the final state, we have a semi-inclusive DIS process,
$e\,p\rightarrow e\,h\,X$.  
Among processes of this kind there are events for which it is possible 
to recognize, in the final hadronic state, particles that bear some identity 
with the original proton, {\textit i.e.}, they are close in rapidity to 
the original proton and carry a significant fraction of its momentum.  
In a particular case, events such as these  
may be characterized  by a large rapidity gap between the products  
of the $\gamma^* p$ hard scattering and the outgoing proton debris  
(Fig.~\ref{fig1}b). 
If those debris are identified with a proton, neutron or any other
baryon close 
related to the original proton, we have the above mentioned {\textit leading  
baryon effect}, $\gamma^{*}\,p\rightarrow X\,N$, which, in analogy with the  
hadron case \cite{Mirian}, could, in principle, be described by Regge  
phenomenology in terms of reggeon and pion exchanges \cite{Szczurek}. 

Furthermore, if the detected baryon is carrying more than 90\% of the
incoming proton momentum and is identified  
with a proton itself (or, equivalently, if a rapidity gap is detected nearby 
the proton fragmentation region), then  
the dominant interaction mechanism is a single diffractive scattering,
$\gamma^{*}\,p\rightarrow X\,p$, in which the virtual photon interacts with  
the proton through  a color singlet exchange with the vacuum quantum numbers,  
which in Regge phenomenology is known as pomeron exchange \cite{Collins}. 
 
With the above statements we just intend to make the point that, speaking 
in terms of theory, diffractive DIS events are part  
of a wider class of interactions, the semi-inclusive DIS processes, within 
which the leading particle effect is found. Thus, if one wants to capture 
the Regge behavior presumably observed by a certain kind of DIS data,
one should take into account all  
available data at once, which, in this case, means to consider simultaneously  
diffractive and leading particle data in the same analysis. This is
the scope of the present paper. 
 
Semi-inclusive processes have been measured by the H1 and ZEUS 
collaborations in the HERA $ep$ colliding machine at DESY, where positrons of  
$27.5\;GeV$ collide with protons of $820\;GeV$. The H1 Collaboration has made  
high-statistic measurements of the 
diffractive structure function $F_{2}^{D}$ in the process $e\,p 
\rightarrow e\,X\,Y$, where $Y$ represents a hadronic system with mass lower 
than $1.6\;GeV$ and rapidity closest to that of the incident proton \cite 
{H1-Diff}. H1 also measured the leading proton and neutron structure 
functions, $F_{2}^{LP}$ and $F_{2}^{LN}$ respectively, in the reaction  
$e\,p \rightarrow e\,N\,X$, where $N$ is the identified nucleon \cite 
{H1-Lead}. The ZEUS Collaboration has measurements of the diffractive 
structure function $F_{2}^{D}$ in the reaction $e\,p \rightarrow 
e\,p\,X$ \cite{Zeus}, and preliminary leading baryon measurements have
also been reported \cite{ZEUS-Lead}.

Now, let us examine these experimental findings through a phenomenological  
gaze. The first attempts to describe them by the Regge formalism were
based on  
the Ingelman and Schlein model \cite{Ingelman} by which diffraction in DIS is  
understood as a two-step process: first the proton emits a pomeron, then the  
pomeron is hard scattered by the virtual photon. In such a view, the pomeron  
is a quasi-particle that carries a fraction $\xi$ of the proton's momentum and 
has its own structure function that could be expressed in terms of $\beta$ and 
$Q^2$ (here $\beta$ plays the role of the Bjorken variable for the pomeron;
see its definition in the next section). Accordingly, the measured structure function  
$F_{2}^{D(4)}(\xi ,t,\beta,Q^{2})$ would be factorized as  
\begin{equation} 
F_{2}^{D(4)}(\xi ,t,\beta,Q^{2})=f_{\tt I\!P}(\xi ,t)\;
F_{2}^{\tt I\!P}(\beta,Q^{2}), 
\label{ism} 
\end{equation} 
where $f_{\tt I\!P}(\xi,t )$ is the flux of the pomeron out of the proton,  
which is a function of $\xi $ and $t$, the squared four-momentum transferred  
at the proton vertex. 
$F_{2}^{\tt I\!P}(\beta,Q^{2})$ represents here the pomeron structure 
function. 
 
Several analyses were made based on Eq.~(\ref{ism}) and on this factorization  
hypothesis, including those performed by the H1 and ZEUS collaborations  
\cite{H1-old,ZEUS-old} (see also \cite{Varios} and references quoted therein). 
In fact, this kind of analyses has been used to establish the pomeron 
intercept $\alpha_{\tt I\!P}$ from the diffractive DIS data. 
  
Although the preliminary experimental results seemed to confirm  
the factorization  
hypothesis  \cite{H1-old,ZEUS-old}, subsequent high-statistic data measured  
in an extended kinematical region by the H1 Collaboration proved that such  
a simple factorized expression is clearly violated \cite{H1-Diff}. 
Since then it has been conjectured \cite{H1-Diff,Golec-Biernat} that secondary 
reggeonic exchanges could 
play an important role in diffractive events, in such a way that the 
structure function could be written as  
\begin{equation} 
F_{2}^{D(4)}(\xi,t,\beta,Q^{2}) = f_{\tt I\!P}(\xi,t)\;
F_{2}^{\tt I\!P}(\beta,Q^{2}) + f_{\tt I\!R}(\xi,t)\;F_{2}^{\tt I\!R}(\beta,Q^{2}),   
\label{strc1} 
\end{equation} 
where $f_{\tt I\!R}(\xi,t)$ is the reggeon flux factor, and  
$F_{2}^{\tt I\!R}(\beta,Q^{2})$ is the reggeon structure
function. Within this  
approach, the change in the diffractive pattern displayed by the H1 data  
could be explained without giving up the idea of Regge factorization for  
each contribution. The H1 Collaboration itself was very successful in  
describing the bulk of the diffractive structure function data with a  
fitting expression akin to Eq.~(\ref{strc1}) (see \cite{H1-Diff}).  
 
In fact, not only the diffractive data, but also the H1 leading proton  
structure function data can be fairly described within the same framework  
as well by just adding up to Eq.~(\ref{strc1}) an extra  
pion contribution as required in such a case (see \cite{H1-Lead}).  
The leading neutron structure function is described by the same scheme, 
but in that case only pion exchange is necessary \cite{H1-Lead}. 
 
Since the leading baryon data were obtained some time after the diffractive 
structure function measurements, these H1 analyses were performed independent 
of each other. 
However, as stated previously, it is our belief that both diffractive and  
leading proton processes should be analyzed together, as two parts of the same 
semi-inclusive process, in the same fashion as in the hadronic case \cite 
{Mirian}. In this way it would be possible to establish more precisely the  
role of the pomeron and the secondary reggeon exchanges, since the 
diffractive  
data are dominated by the former and has the latter only as a background,  
while the reverse is true for the leading proton data.  
Therefore, in this work we consider these data sets as 
complementary ones, {\textit i.e.}, our basic assumption 
is that the diffractive and 
leading proton structure functions are parts of one and the same 
semi-inclusive proton structure function , which can be expressed in a way  
similar to Eq.~(\ref{strc1}). Throughout this
work we will use the notation $F_{2}^{SI}$ for the semi-inclusive 
proton structure function, when referring to the diffractive and leading 
proton structure function data together.

The purpose of this paper is to reach a better understanding about the  
role of the pomeron and reggeon contributions in the interface between the  
diffractive and non-diffractive regimes through a global fit of the  
proton structure function obtained from H1 semi-inclusive DIS  
data (the ZEUS data were not employed in the fitting procedure, but
their diffractive structure function measurements were used for
checking our final results). In Sec.\ref{Kinematics},
we define the kinematical variables and cross 
sections while our fitting procedure is presented in Sec.\ref{Model}. 
In Sec.\ref{Results}, we present our fit results and a preliminary discussion,
while a procedure to compare diffractive and leading proton data 
is described in Sec.\ref{Bringing}.
Our main conclusions are summarized in Sec.\ref{Conclusions}.

\section{\label{Kinematics}Kinematics and Cross Sections} 
 
The usual variables employed to describe $ep$ DIS are depicted in  
Fig.~\ref{fig1}a. One can define the squared energy in the $ep$ center
of mass system (CMS) in terms  
of the four-momenta $P$ and $k$, referring respectively to the incoming  
proton and electron (or positron), as 
\begin{equation} 
s \; = \; (P + k )^{2}  \label{S} 
\end{equation} 
\noindent and the squared energy in the $\gamma^{*} p$ CMS as  
\begin{equation} 
W^{2} \; = \; (P+q)^{2}.  \label{W} 
\end{equation} 
The photon virtuality $Q^{2}$, the Bjorken $x$ and the variable $y$ 
are given by 
\begin{eqnarray*} 
q^{2} & = & -Q^{2} = (k - k^{,})^{2}, \\ 
x & = & {\frac{Q^{2}}{2\ P\cdot q}} = \frac{Q^2}{W^2 + Q^2 - m_p^2}, \\
y & = & {\frac{P \cdot q }{P \cdot k}}. 
\end{eqnarray*} 
\noindent If we ignore the proton mass, we have the following 
relations among these variables:  
\begin{equation} 
Q^{2} = x\ y\ s  
\end{equation} 
and  
\begin{equation} 
W^{2} = Q^{2}\,\frac{(1-x)}{x} 
\simeq \frac{Q^{2}}{x},   
\label{QW2} 
\end{equation} 
being that $x << 1$ has been assumed in the latter expression. 
 
For the case presented in Fig.~\ref{fig1}b, where a baryon with 
four-momentum $P^{\prime}$ is detected in the final state, we can also 
define the variables 
\begin{eqnarray} 
t & = & (P - P^{\prime})^{2}, \\ 
\nonumber \\ 
\xi & = & {\frac{Q^{2}+M_{X}^{2}-t }{Q^{2}+W^{2}}}, \\ 
\nonumber \\ 
\beta & = & {\frac{Q^{2} }{Q^{2} + M_{X}^{2} -t}} = {\frac{x }{\xi}}, 
\label{beta} 
\end{eqnarray} 
where the $\beta$ variable represents the fraction of momentum carried by a 
struck parton in the pomeron (if a pomeron exchange model is assumed).  

Also, for leading baryons, it is usual to 
describe the data in terms of the fraction of momentum carried by 
the outgoing proton, $z = P^{\prime}/P$,
where $z$ is connected with $\xi$ by
\begin{equation}
z = 1 - \xi.
\end{equation}

The differential cross section for a semi-inclusive DIS process 
giving rise to leading baryon behavior is 
written as  
\begin{equation} 
{\frac{d^{3}\sigma }{dx\ dQ^{2}\ d z }}={\frac{4\pi\ \alpha _{em}^{2}} 
{x\ Q^{4}}} 
\left[1-y+{\frac{y^{2}}{2(1+R)}}\right] \;F_{2}^{LB(3)}(z ,x,Q^{2}). 
\label{DIS-x} 
\end{equation} 
In the case of diffractive events, such a cross section is often expressed 
in terms of the $\beta$ and $\xi$ variables, 
\begin{equation} 
\frac{d^{3}\sigma}{d\beta\ dQ^{2}\ d\xi}= 
\frac{4\pi\ \alpha _{em}^{2}}{\beta\ Q^{4}} 
\left[1 - y + \frac{y^{2}}{2(1+R)}\right] 
\;F_{2}^{D(3)}(\xi,\beta,Q^{2}). 
\label{DIS-beta} 
\end{equation} 
Here $R=\sigma_{L}/\sigma _{T}$ is the ratio between the cross sections for 
longitudinally and transversely polarized virtual photons. Under certain 
conditions, it is possible to assume $R\approx 0$ and thus the experimental 
behavior of the cross sections (\ref{DIS-x}) and (\ref{DIS-beta}) is 
expressed in terms of the structure functions $F_{2}^{LB(3)}(z,x,Q^{2})$  
and $F_{2}^{D(3)}(\xi,\beta,Q^{2})$. Specifically for the H1 diffractive  
data, such assumption was applied for those data with $y<0.45$ \cite{H1-Diff}. 
 
Thus, our analysis is directed to study the behavior of both 
$F_{2}^{LB(3)}(z,x,Q^{2})$ and $F_{2}^{D(3)}(\xi,\beta,Q^{2})$ data. We 
notice that these data are already integrated over the $t$-range 
corresponding to their respective experiments. In order to compare these 
data among themselves it is necessary to explicitly introduce the 
$t$-dependence on the structure functions. We discuss that issue in
details in Sec. \ref{Bringing}.

\newpage
\section{\label{Model}Model, Parameters and Fitting Procedure}

In the present study we have used the diffractive structure function 
data $F_{2}^{D}$ obtained by the H1 Collaboration \cite{H1-Diff}, together
with their measurements of the leading baryon structure functions 
$F_{2}^{LP}$ for protons and $F_{2}^{LN}$ for neutrons \cite{H1-Lead},
in the same analysis.
The $F_{2}^{D}$ data cover the kinematical ranges: 
\begin{eqnarray*}
1.2 \cdot 10^{-4} \; < & x & < \; 2.37 \cdot 10^{-2}, \\
4.5 \; < & Q^{2} & < \; 75 \; GeV^{2}, \\
0.04 \; < & \beta & < \; 0.9.
\end{eqnarray*}
while, for the leading baryon $F_{2}^{LB}$ measurements, the 
covered kinematical region are: 
\begin{eqnarray*}
10^{-4} \; < & x & < \; 3.3 \cdot 10^{-3}, \\
2.5 \; < & Q^{2} & < \; 28.6\; GeV^{2}, \\
3.7 \cdot 10^{-4} < & \beta & < 2.7 \cdot 10^{-2}.
\end{eqnarray*}

We notice that, although these data sets are overlapping in terms of
$x$ and $Q^2$
ranges, they are complementary in terms of the $\beta$, the
Bjorken variable for the presumable pomeron constituents.

As stated before, the H1 diffractive structure function, 
$F_{2}^{D(3)}(\xi,\beta,Q^{2})$, can be written as
a combination of two Regge exchanges with the quantum numbers of
the vacuum, the pomeron and the reggeon ones \cite{H1-Diff}.
The most general expression for such a diffractive structure function reads

\begin{equation}
F_{2}^{D(3)}(\xi,\beta,Q^{2}) = g_{\tt I\!P}(\xi)\;
F_{2}^{\tt I\!P}(\beta,Q^{2}) +  g_{\tt I\!R}(\xi)\;F_{2}^{\tt I\!R}(\beta,Q^{2}) +
g_{\tt I}(\xi)\;F_{2}^{\tt I}(\beta,Q^{2}).
\label{F2D3}
\end{equation}

Here, functions $g_{\tt I\!P}(\xi)$ and $g_{\tt I\!R}(\xi)$
represent, respectively, the pomeron and reggeon flux factors 
integrated over $t$, while $F_{2}^{\tt I\!P}(\beta,Q^{2})$ and 
$F_{2}^{\tt I\!R}(\beta,Q^{2})$ are the pomeron and reggeon structure
functions. The last term on the right-hand-side of Eq.~(\ref{F2D3}),
$g_{\tt I}(\xi)\;F_{2}^{\tt I}(\beta,Q^{2})$, accounts for a
possible interference effect between the pomeron and reggeon exchanges.

The fluxes are taken from the Regge phenomenology of hadronic soft
diffraction, and are written as
 
\begin{equation}
g_{\tt I\!P}(\xi) = \xi ^{1-2\alpha _{\tt I\!P}^{0}}\;
\int_{|t_{min}|}^{|t_{max}|} e^{-(\alpha _{\tt I\!P}^{,}
ln{\xi })\,t}\;F_{1}^{2}(t)\;dt
\label{pomeron_flux}
\end{equation}
and
\begin{equation}
g_{\tt I\!R}(\xi) = \xi ^{1-2\alpha _{\tt I\!R}^{0}}\;
\int_{|t_{min}|}^{|t_{max}|}\,e^{({b_{\tt I\!R}^{0}-
\alpha_{\tt I\!R}^{,}ln{\xi }})t}\;dt,
\label{reggeon_flux}
\end{equation}

\noindent where $|t_{min}|$ and $|t_{max}|$ are the minimum and 
maximum absolute $t$ values of the data for each experiment.
In these expressions, the parameters $\alpha _{\tt I\!P}^{0}$, 
$\alpha_{\tt I\!R}^{0}$ and
$\alpha_{\tt I\!P}^{,}$, $\alpha _{\tt I\!R}^{,}$ are, respectively, the
intercept and slope of the pomeron and reggeon linear trajectories, that is
\begin{equation}
\alpha _{\tt I\!P}(t) =  \alpha _{\tt I\!P}^{0}+\alpha _{\tt
I\!P}^{,}\;t \ \ \ \ \ \ \ \ {\rm and}\ \ \ \ \ \ \ \ \ 
\alpha _{\tt I\!R}(t)  =  \alpha _{\tt I\!R}^{0}+\alpha _{\tt I\!R}^{,}\;t,
\end{equation}
and $F_{1}(t)$ in Eq.~(\ref{pomeron_flux}) is the Dirac form factor given by 
\begin{equation}
F_{1}(t)={\frac{4m_{p}^{2}-0.28t}{4m_{p}^{2}-t}} \bigg({\frac{1}
{1-t/0.71}}\bigg)^{2}.
\end{equation}

The interference term $g_{\tt I}(\xi)\;F_{2}^{\tt I}(\beta,Q^{2})$
is related to the pomeron and reggeon fluxes and structure functions
by 
\begin{equation}
F_{2}^{\tt I}(\beta,Q^{2}) = \sqrt{F_{2}^{\tt I\!P}(\beta,Q^{2}) \;
F_{2}^{\tt I\!R}(\beta,Q^{2})}
\label{F2interference}
\end{equation}
and
\begin{equation}
g_{\tt I}(\xi) = 2\;I\;
\int_{|t_{min}|}^{|t_{max}|}
\cos\{\frac{\pi}{2}[\alpha_{\tt I\!P}(t) - \alpha_{\tt I\!R}(t)]\} \;
\sqrt{e^{b_{\tt I\!R}\,t}\;F_{1}^{2}(t)}\;
\xi^{1 - \alpha_{\tt I\!P}(t) - \alpha_{\tt I\!R}(t)}\;dt.
\label{interference_flux}
\end{equation}

The expression above is quite similar to the one used by the H1
Collaboration to
account for interference contribution in their diffractive
structure function analysis \cite{H1-Diff}.
Following their procedure, we
introduced a free parameter $I$ to account for the degree of
interference between the pomeron and reggeon exchanges. Such
a parameter is allowed to vary from 0 to 1.

Here we mostly intend to explore the connection between the
diffractive and leading proton regimes, although the available data
are quite separated in terms $\beta$. Therefore, we
need a general functional form for the pomeron structure function 
that could be able to consider both the
low $\beta$ (leading proton) and high $\beta$ (diffractive)
regimes. In order to do that, we choose for the pomeron a functional
form based on the
same phenomenological parameterization as used in the H1 QCD
analysis of the diffractive structure function \cite{H1-Diff}, where
a quark flavor singlet distribution $\beta S_{q}(\beta,Q^{2}) =
u+\bar{u} + d + \bar{d} + s + \bar{s}$ and a gluon distribution
$\beta G(\beta,Q^{2})$ are parameterized in terms of the coefficients
$C_{j}^{(S)}$ and $C_{j}^{(G)}$, according to:

\begin{eqnarray}
\beta S(\beta,Q^{2}=Q_{0}^{2}) & = &
\left[ \sum_{j=1}^{n} C_{j}^{(S)} \cdot P_{j} (2\beta-1)\right]^{2}
\cdot exp{(\frac{a}{\beta-1})} \nonumber \\
\beta G(\beta,Q^{2}=Q_{0}^{2}) & = &
\left[ \sum_{j=1}^{n} C_{j}^{(G)} \cdot P_{j} (2\beta-1)\right]^{2}
\cdot exp{(\frac{a}{\beta-1})}.
\label{F2pomeron}
\end{eqnarray}
where $P_{j}(\zeta)$ is the $j^{th}$ member in a set of Chebyshev
polynomials, with $P_{1}=1$, $P_{2}=\zeta$ and $P_{j+1}(\zeta)=
2\zeta P_{j}(\zeta)-P_{j-1}(\zeta)$. We have summed these terms up to
$n=3$ and set
$Q_{0}=2\,GeV^{2}$, in order to contemplate the $Q^{2}$ range of both
diffractive and leading proton data. Following H1, we also set $a=0.01$.
Therefore, Eq. (\ref{F2pomeron}) has 6 parameters to be fixed by the
fit. 

Since it is not possible
to totally separate the pomeron structure function from its flux
factor, the parameters $C_{j}^{(S)}$ above also set the overall normalization
of the pomeron contribution. The gluon and quark distributions
above are evolved in leading order (LO) and next-to-leading order
(NLO) by using the QCDNUM16 package \cite{QCDNUM},
and the final pomeron structure function is written in terms of
the singlet quark distribution as
\begin{equation}
F_{2}^{\tt I\!P} (\beta,Q^{2}) = <e^{2}>
(u+\bar{u} + d + \bar{d} + s + \bar{s})
\end{equation}
where $<e^{2}>$ is the average charge of the distribution, and for
three flavors $<e^{2}> = 2/9$.

For the reggeon, we assume the hypothesis of a direct relation between
the reggeon structure function and the pion structure function by using
\begin{equation}
F_{2}^{\tt I\!R} (\beta,Q^{2}) = N_{\tt I\!R} \, F_{2}^{\pi}(\beta,Q^{2}),
\label{F2reggeon}
\end{equation}
\noindent where $N_{\tt I\!R}$ is a free normalization parameter,
and for the pion structure function we choose the LO GRV
parameterization \cite{GRV}. Such  a choice is supported by the good
description it provided for the H1 leading baryon data
\cite{H1-Lead}.

In fact, the identification of the reggeon structure function with the pion
one is not new, and some authors already have applied it to the
analysis of the H1 diffractive structure function data \cite{Royon}.

Specifically for our case, we also choose to identify the reggeon
exchange explicitly with the $f_2$
family of resonances, which has the right quantum numbers for the
processes analyzed here and is characterized by its high 
intercept, $\alpha^{0}_{\tt I\!R} \approx 0.68$ \cite{Dino1}.

For the leading proton structure function,
$F_{2}^{LP(3)}(\xi,\beta,Q^{2})$, besides the pomeron and reggeon
contributions, the pion exchange also plays a major role. In fact,
the pion contribution is known to have an important role in hadronic
leading proton \cite{Mirian} and seems to work as an effective background for
$\bar{p}p$ diffractive reactions at small $t$ \cite{Dino2}, besides
its role in DIS \cite{H1-Lead}. Indeed, pion exchange has a well known
phenomenological behavior, so we took the pion flux factor straight out of
the literature as being
\begin{equation}
f_{\pi }(\xi ,t)={\frac{g_{pp}}{4\pi }}{\frac{1}{4\pi }}{\frac{|t|}
{(t-0.02)^{2}}}\xi ^{1-2\alpha _{\pi }(t)},
\label{pion_flux}
\end{equation}
where $g_{pp}/4\pi =13.6$ is the coupling constant for $pp\rightarrow pX$.
Note that for the inclusive neutron
production, $pp\rightarrow nX$, there is an extra factor $2$
in the coupling constant due to the Clebsh-Gordan coefficient for such a
process.

For the pion structure function, $F_{2}^{\pi}(\beta,Q^{2})$,
we took the LO GRV \cite{GRV} parameterization. With the flux
above and the GRV structure function, we were successful in describing the
DIS leading neutron data without any free parameter.

The expression for the leading proton structure function then
reads
\begin{equation}
F_{2}^{LP(3)}(\xi,\beta,Q^{2}) = g_{\tt I\!P}(\xi)\;
F_{2}^{\tt I\!P}(\beta,Q^{2}) +
g_{\tt I\!R}(\xi)\;F_{2}^{\tt I\!R}(\beta,Q^{2}) +
g_{\pi}(\xi)\;F^{\pi}(x,Q^{2}).
\label{F2LP3}
\end{equation}

As we said at the beginning, our main assumption is that
the diffractive and leading proton structure function are components
of one and the same semi-inclusive (SI) structure function, which combines
the contribution from both Eq.~(\ref{F2D3}) and Eq.~(\ref{F2LP3}) in a
single expression that reads

\begin{eqnarray}
F_{2}^{SI(3)}(\xi,\beta,Q^{2}) = g_{\tt I\!P}(\xi)\;
F_{2}^{\tt I\!P}(\beta,Q^{2}) & + &
g_{\tt I\!R}(\xi)\;F_{2}^{\tt I\!R}(\beta,Q^{2}) \nonumber \\ & + &
g_{\pi}(\xi)\;F^{\pi}(\beta,Q^{2}) + g_{I}(\xi)\;F_{2}^{I}(\beta,Q^{2}).
\label{F2SI}
\end{eqnarray}

It should be noted that in the equation above, the pion contribution is
significant only for $\xi \geq 0.1$, therefore for the diffractive regime,
$\xi \leq 0.05$,  Eq.~(\ref{F2SI}) reduces to Eq.~(\ref{F2D3}), where no pion
exchange is considered.

In overall, we dealing with a maximum of 8 free parameters to be fixed
by the fitting procedure.
These parameters come from the pomeron structure function,
Eq.~(\ref{F2pomeron}) (6 parameters), reggeon normalization,
Eq.~(\ref{F2reggeon}) (1 parameter), and the interference contribution,
Eq.~(\ref{F2interference}) (1 parameter). As mentioned before, the
pion contribution (flux factor, Eq.~(\ref{pion_flux}), and structure
function, given by the GRV parameterization \cite{GRV})
is totally fixed by the standard phenomenology having no free parameter left.

The other parameters, such as the pomeron and reggeon trajectories (intercept
and slopes), the slope of the reggeon $t$-dependence and the $a$
parameter from Eq.~(\ref{F2pomeron} were kept fixed by their values from
the literature, since they are quite well established. In Table \ref{table1} 
we present the values used for these parameters throughout this paper.

It should be mentioned that we excluded from
the fit all data lying the the resonance region ($M_{X}^{2} \leq
2 \;GeV^{2}$) and/or with $y \geq 0.45$. That leave us with a
total of 170 diffractive structure function data and 48 leading
proton structure function data, with adds to a total of 218 data.

\begin{table}[ht]
\caption{\label{table1}Values used for the parameters that were kept
fixed during the fitting procedure.}
\begin{ruledtabular}
\begin{tabular}{ccccccc}
Parameters: & $\alpha_{\tt I\!P}$ & $\alpha_{\tt I\!P}^{,}$ & 
$\alpha_{\tt I\!R}$ & $\alpha_{\tt I\!R}^{,}$ & $b_{\tt I\!R}^{0}$ &
$a$ \\ \hline
Values: & $1.2$ & $0.25\,GeV^{-2}$ & $0.68$ & $0.9\,GeV^{-2}$ &
$2.0\,GeV^{-2}$ & $0.001$  \\
\end{tabular}
\end{ruledtabular}
\end{table}

\section{\label{Results}Results and Discussion}

In Table~\ref{TableFit}, we present the results of our first three fits.
Fit~1 represents the results of our global LO analysis
of diffractive and leading proton structure function data, using 
Eq.~(\ref{F2SI}) with no interference term included ($I=0$).
Since we are dealing with two different sets of data, we added
the statistic and systematic errors in quadrature. A $\chi^{2} /
d.o.f.$ of $1.277$ was obtained.

Fit~2 corresponds to the results of a global NLO analysis of the 
diffractive and leading proton structure function data, using 
Eq.~(\ref{F2SI}) as the fitting equation, again with no interference term 
included ($I=0$). Although some of the parameters have significantly
changed in comparison to Fit~1, the final result provided a 
$\chi^{2}/ d.o.f. = 1.276$ which is basically the same as the one from the global 
LO fit.

Fit~3 corresponds to a fit of Eq.~(\ref{F2D3}) to the diffractive structure
function data only. The final $\chi^{2} / d.o.f.$ obtained, with only
statistical errors included, was $\chi^{2} / d.o.f. = 1.106$.
Although, in this case, the interference component was left free, it
was ruled out by the fit. An observation to be made at this point is that  
one must be careful when comparing this $\chi^{2} / d.o.f.$ result with 
the one from the H1 QCD analysis of the same set of data
\cite{H1-Diff}, since our sample includes two sets of data that
where not taken into account in the H1 analysis (those for
$Q^{2}=45\,GeV^{2}$ and $Q^{2}=75\,GeV^{2}$ at $\beta = 0.9$).
That gives us a total of $170$ data, whereas H1 has only $161$.
Our choice for the reggeon intercept has also some effect in
improving the final $\chi^{2}$ result.

\begin{table*}[ht]
\caption{\label{TableFit}Parameters obtained from the fits
to diffractive and leading proton structure function data. For this
results, the interference parameter was turned off ($I=0$). 
The individual contribution to the $\chi^{2}$
coming from the diffractive (Diff.) and leading proton (LP) data
are also presented, with their relative weight in the
total $\chi^{2}$ (in $\%$) presented in parenthesis. All errors are quoted
as obtained from MINUIT.}
\begin{ruledtabular}
\begin{tabular}{cccc} 
\hline
Parameters & Fit 1 - Global LO & Fit 2 - Global NLO & Fit 3
-Diffractive NLO            \\ \hline
$C_{1}^{(S)}$  & $0.111 \pm 0.031$ & $0.116 \pm 0.017$ & $0.147 \pm 0.040$  \\
$C_{2}^{(S)}$  & $0.076 \pm 0.034$ & $0.169 \pm 0.029$ & $0.182 \pm 0.053$  \\
$C_{3}^{(S)}$  & $0.156 \pm 0.034$ & $0.181 \pm 0.035$ & $0.065 \pm 0.038$  \\
$C_{1}^{(G)}$  & $1.110 \pm 0.056$ & $0.710 \pm 0.052$ & $0.704 \pm 0.095$  \\
$C_{2}^{(G)}$  & $0.817 \pm 0.071$ & $1.350 \pm 0.053$ & $1.079 \pm 0.167$  \\
$C_{3}^{(G)}$  & $0.284 \pm 0.097$ & $0.633 \pm 0.168$ & $0.306 \pm 0.180$  \\
$N_{\tt I\!R}$    & $2.048 \pm 0.124$ & $2.058 \pm 0.123$ & $7.25 \pm 0.55$ \\
$\chi^{2}$ (Diff.) & $202.60 \,\, (75\%)$  & $199.73 \,\, (74\%)$  &
$180.23 \, \, (100\%)$ \\
$\chi^{2}$ (LP)    & $66.90 \,\, (25\%)$   & $69.48 \,\, (26\%)$  & - \\
$\chi^{2}/d.o.f.$  & $269.50/(218-7)$ & $269.21/(218-7)$  &
$180.23/(170-7)$ \\
\end{tabular}
\end{ruledtabular}
\end{table*}

\begin{table}[ht]
\caption{\label{TableFit2}Parameters obtained from the global fits 
to diffractive and leading proton structure function data. For this
results, the interference parameter was set free.
The individual contribution to the $\chi^{2}$
coming from the diffractive (Diff.) and leading proton (LP) data
are also presented, together with the total $\chi^{2}$.
All errors are quoted as obtained from MINUIT.}
\begin{ruledtabular}
\begin{tabular}{ccc}
Parameters & Fit 4 - Global LO & Fit 5 - Global NLO      \\ \hline
$C_{1}^{(S)}$  & $0.166 \pm 0.024$ & $0.121 \pm 0.019$   \\
$C_{2}^{(S)}$  & $0.056 \pm 0.029$ & $0.167 \pm 0.023$   \\
$C_{3}^{(S)}$  & $0.083 \pm 0.037$ & $0.174 \pm 0.037$   \\
$C_{1}^{(G)}$  & $0.874 \pm 0.073$ & $0.711 \pm 0.049$   \\
$C_{2}^{(G)}$  & $0.854 \pm 0.142$ & $1.180 \pm 0.063$   \\
$C_{3}^{(G)}$  & $0.124 \pm 0.108$ & $0.578 \pm 0.014$   \\
$N_{\tt I\!R}$ & $1.396 \pm 0.119$ & $1.259 \pm 0.112$   \\
$I$            & $1.0 \pm 0.694$   & $1.0 \pm 0.0805$    \\
$\chi^{2}$ (Diff.) & $176.90 \,\, (73\%)$ & $169.78 \,\, (68\%)$ \\
$\chi^{2}$ (LP) & $66.30 \,\, (27\%)$  & $78.77 \,\, (32\%)$     \\
$\chi^{2}/d.o.f.$ & $243.20/(218-8)$ & $248.55/(218-8)$  \\
\end{tabular}
\end{ruledtabular}
\end{table}

Table \ref{TableFit2} presents the results of global fits when
the interference parameter $I$ set free. It was bounded to vary
in the interval $0 \le I \le 1$, but, as can be seen, in both fits it 
assumed the
maximum upper value. Comparing these results respectively to Fits 1 and 2, the
$\chi^{2}/d.o.f.$ improved a little in both
the LO fit ($\chi^{2}/d.o.f.=1.16$) and the NLO fit ($\chi^{2}/d.o.f.=1.18$).

Table \ref{TableFit} and \ref{TableFit2} also present the individual
contributions to the $\chi^{2}$ coming from the diffractive and
leading proton data. For three of our global fits, we have a diffractive
contribution around $74\%$, with the leading proton one around $26\%$.
The only departure from these values comes from the global NLO fit
with the interference parameter $I$ set free (Fit 5). For that we have
the diffractive data contributing with $68\%$ and the leading proton
data with $32\%$.
It is worth to remember that, for the global fits, our
data sample is composed of 218 data, 170 coming from diffractive
and 48 from leading proton structure function. Therefore, the
diffractive data corresponds to $78\%$ of our global data set, and the
leading proton data to $22\%$. 

\begin{figure}[ht]
\includegraphics[height=14cm]{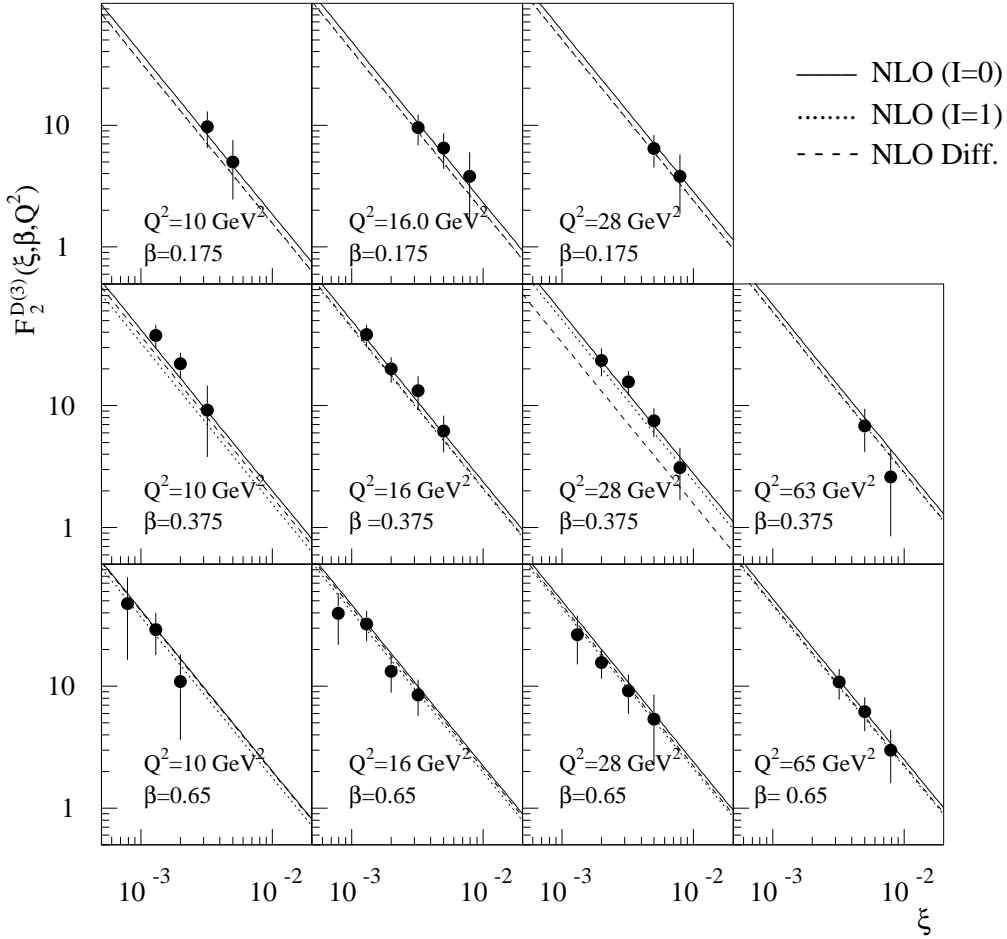}
\caption{\label{fig2}Diffractive structure function data $F_{2}^{D(3)}$ from
Zeus Collab. \cite{ZEUS-old}, together with the results for the pomeron
structure function extracted from Fit 2 (solid line), Fit 3 (dashed
line) and Fit 5 (dotted line).}
\end{figure}

In order to test the parameterization of the pomeron structure
function, we compare some of our results for $F_{2}^{\tt
I\!P}(\xi,\beta,Q^{2})$, Eq.~(\ref{F2pomeron}), with the independent
measurement of $F_{2}^{D(3)}(\xi,\beta,Q^{2})$ by the ZEUS Collaboration
\cite{ZEUS-old}, where no sign of secondary exchanges was found.  
As shown in Fig.~\ref{fig2}, all of the three fits exhibited are in
good agreement with
the data (which were not used in the fitting procedure), indicating
that the pomeron contribution has been fairly accounted.

\begin{figure}[ht]
\includegraphics[height=16cm]{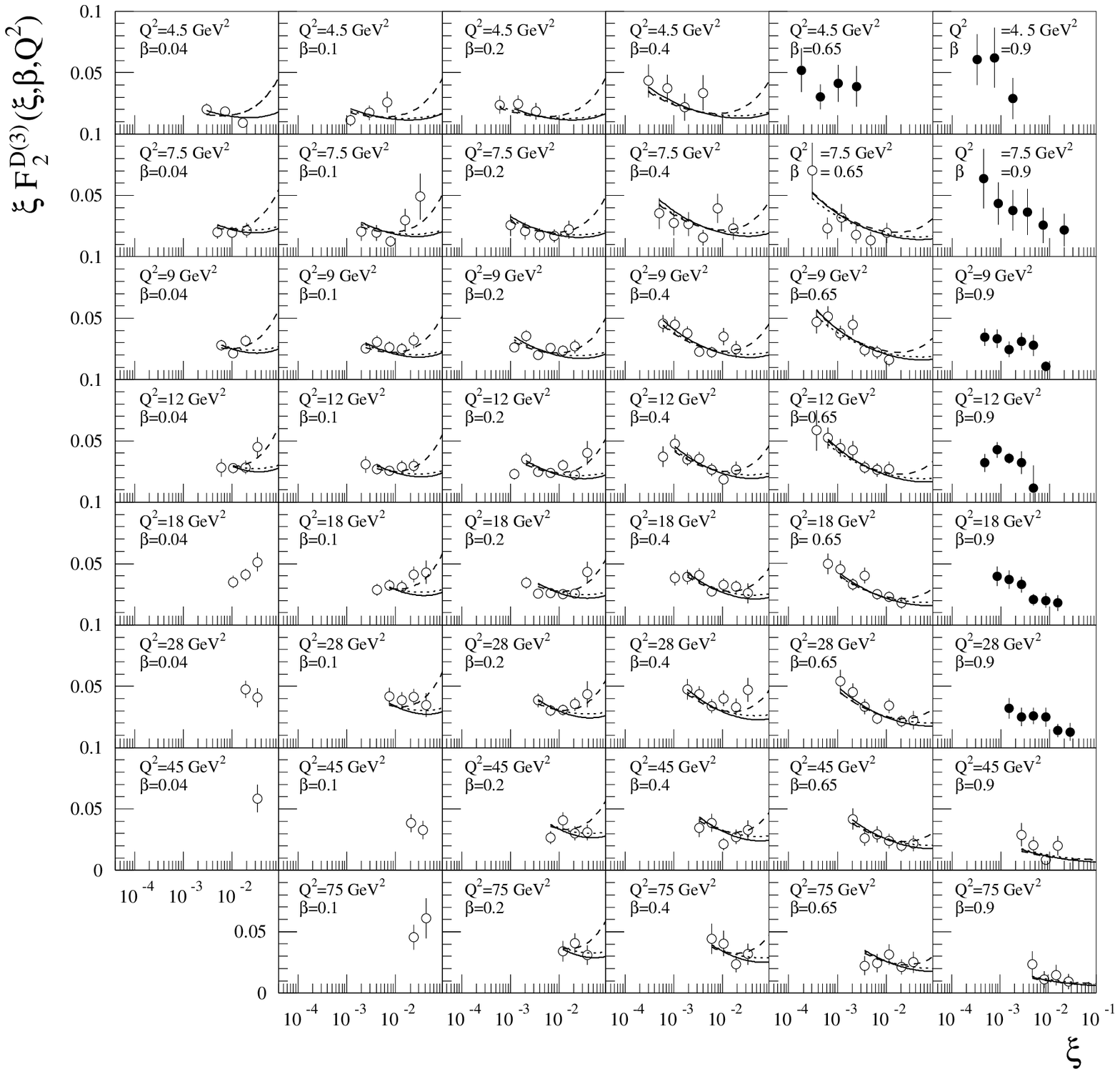}
\caption{\label{fig3}Plot of the H1 diffractive structure function data
$\xi \; F_{2}^{D(3)}$ for fixed $\beta$ and $Q^{2}$.
The curves represent the best fit resultant from our global NLO
Fit 2 (solid line) and Fit 5 (dotted line). We also show the
diffractive NLO Fit 3 (dashed line). Those data points lying in the
resonance region, $M_{X}^2 < 2\;GeV^2$, are displayed as black circles.}
\end{figure}

In Fig.~\ref{fig3}, we plotted the diffractive structure function data from H1
Collaboration in comparison with the results of the same three fits
shown in Fig~1. As can be seen, the agreement among the three fits
is quite good at small $\xi$, but as $\xi$ increases Fit~3 grows
faster than the other two. The difference between Fit~1 and Fit~2 (not
shown in the figure) is quite
small over the entire diffractive range of $\xi$, which is expected
since both fits give close values for the $\chi^{2}$.

\begin{figure}[ht]
\includegraphics[height=16cm]{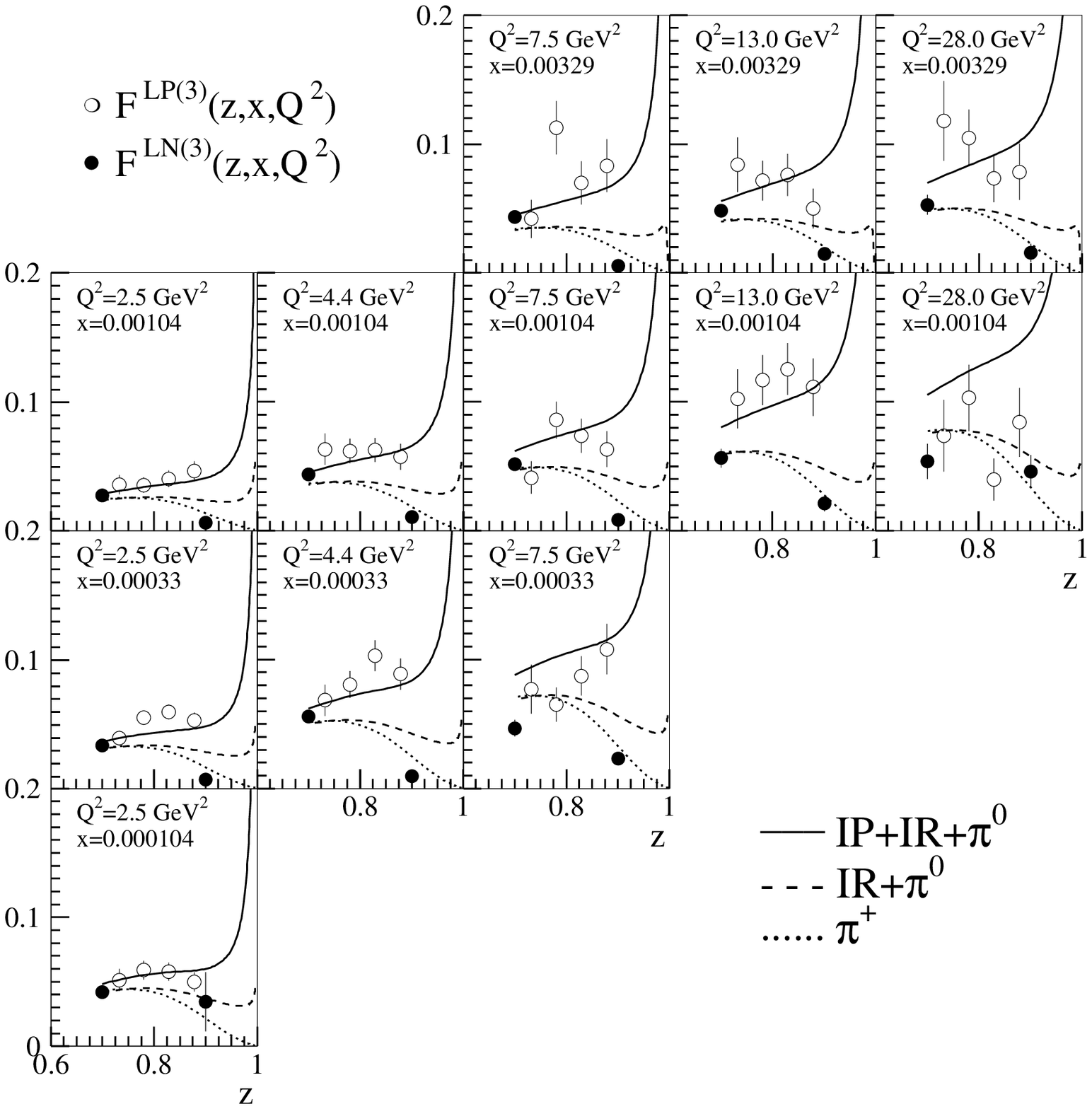}
\caption{\label{fig4}Plots of the leading proton structure function 
data $F_{2}^{LP}$
(for fixed $x$ and $Q^{2}$) vs. $z=1-\xi$, with result of the global NLO fit
with no interference (Fit 2). The leading
neutron structure function data $F_{2}^{LN}$ is also shown, together
with the prediction coming from the product of the standard pion flux,
Eq.(\ref{pion_flux}), and the GRV parameterization of the pion
structure function, $F_{2}^{\pi}(\beta,Q^{2})$ \protect\cite{GRV}.}
\end{figure}

Fig.~\ref{fig4} shows the $F_2^{LP}$ data from H1 Collaboration
together with the results from Fit~2 to illustrate the description of
the leading particle behavior. The leading neutron
data, from the same experiment, are also included (these data can be
described assuming pion exchange as the only contribution 
for the reaction and so were not employed in the fitting procedure).

After showing all of these results, some comments are in order. Firstly,
from Fits 1 and 2, we see that applying LO or NLO evolution equations 
produce basically the same result in terms of $\chi^2$, although, as
expected, some parameters suffer a little change (the same can be said
about Fits 4 and 5). We remind that these parameters reflect the quark
and gluon content of the pomeron as obtained from different scenarios.

The comparison between Fit 2 (global) and Fit 3 (only
diffractive data) present much more remarkable effects. Not only the
parameters change, but in the latter case there is a strong
enhancement of the secondary contribution. However, this is a
suspicious effect since the diffractive data are quite limited in
terms of the $\xi$ variable and secondary reggeon contribution are
supposed to play an important role only for $\xi \geq 0.1$ (see more
comments about this aspect in the next section).

When we perform the global fit, but leaving the interference term
completely free to be established by the $\chi^2$ minimization, it
assumes its maximum value (Fits 4 and 5). Again it is the case of
asking whether this outcome reflects a reliable physical effect or 
is just a fitting artifact. Answering this question is beyond the
scope of this paper, but we have strong evidences indicating that the
introduction of the interference term makes the corresponding structure
functions inadequate to describe the results of diffractive photo- and 
eletroproduction of dijets by both H1 and ZEUS collaborations. On the
other hand, diffractive structure function obtained without
interference effects allow a very good description of both dijet
production processes \cite{Altem}.

\section{\label{Bringing}Bringing diffractive and leading proton
structure functions together}

\begin{figure*}[ht]
\includegraphics[height=12cm]{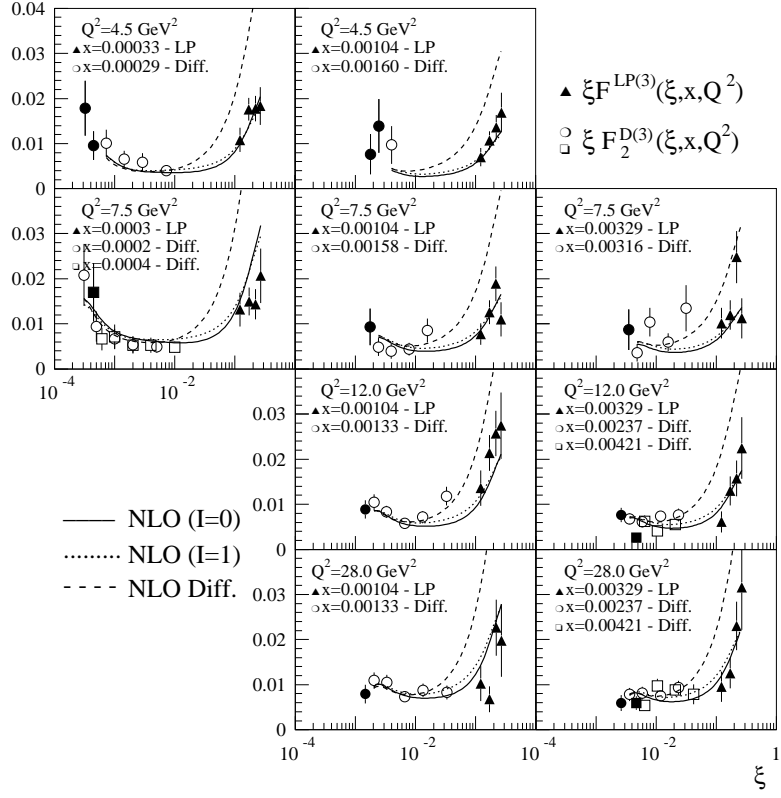}
\caption{\label{fig5}Diffractive (Diff, white circles and white squares) and leading
proton (LP, black triangles) structure function data vs. $\xi$, for fixed
$x$ and $Q^{2}$.
The figure combines, in each plot, the diffractive and leading proton
data with the same or close values of $x$ and $Q^{2}$.
The quoted $Q^{2}$ values are
those of the diffractive data, for which the correspondent
leading proton values
are $Q^{2} = 4.4, \;\;7.5, \;\; 13.3$ and $28.6\;GeV^{2}$. 
The black circles and black squares represent data with $M_{X}^2 < 2\;GeV^2$.
For a matter of presentation, every leading proton data
was multiplied by a scale factor, to compensate for their shorter $t$
range compared to the diffractive $F_{2}^{D(3)}$ measurements, as
explained in the text. The plotted curves represent our global best
NLO fits, with no interference (solid line) and maximum interference
(dotted line), and the diffractive NLO fit alone (dashed line).}
\end{figure*}

Now, some words are needed to explain how we handled together both the 
sets of data displayed in Fig.~\ref{fig5}, since it is the 
central piece of our study. In that figure, we bring together the
diffractive and leading proton data and compare the results of our
three NLO fits to this combined set of semi-inclusive data.

Here, we are mostly interested in analyzing the behavior
of these data in terms of $\xi$. Since the $\beta $ range
for the diffractive and leading proton data are very
distinct, the usual procedure of plotting together data with the same 
values of  $\beta$ and $Q^{2}$ would not be the best choice. 
There is, however, a large overlap of these two sets in terms of 
the variables $x$ and $Q^{2}$. Thus, we choose to combine the data 
in groups with the same (or as close as possible) values of $x$ and $Q^{2}$. 
That is a more
proper way to show that the difference between the diffractive and the 
leading proton regime is due to the $\xi$ region where the
semi-inclusive process $e p \rightarrow e p X$ is measured, according to 
our assumption that both sets of data can be embraced by the same
semi-inclusive structure function.

Still a problem remains. Besides the different $\beta$ range, both the
diffractive and leading proton structure functions were measured
at different $t$ intervals. The diffractive data were measured for
the interval $|t_{min}| < |t| < 1\,GeV^{2}$, whereas the leading
proton ones where measured for the interval $|t_{min}| < |t| < |t_{0}|$,
where
\begin{eqnarray}
t_{min} & = & - \frac{m_{p}^{2}\ \xi^{2}}{(1-\xi)}, \\
t_0 & = & -\frac{p_{T, max}^2}{(1 - \xi)} + t_{min},
\end{eqnarray}
\noindent with  $p_{T, max} = 0.2 \; {\rm GeV}$.
Since this last interval corresponds to a  range smaller than the
diffractive one and since the phenomenological $t$ dependence
coming from the diffractive region seems to be well established
for both hadronic and DIS events, in Fig.~\ref{fig5} we scaled down the
diffractive structure function data in order to make them comparable 
to the leading proton data.

It should be noticed that such a correction is intended only
as a visualization device. In our whole fitting analysis, we took
the data at their correct measured $t$ intervals.

In order to make such a correction as independent of our own analysis 
as possible,
we choose to proceed by the following way. A fit of Eq.~(\ref{F2D3}) to the
diffractive structure function data was performed, with the
interference parameter $I$ set to zero (no interference).
The fluxes were those given by Eq.~(\ref{pomeron_flux}) and
Eq.~(\ref{reggeon_flux}), with the pomeron and reggeon
intercept kept fixed with those values obtained from the H1 analysis
\cite{H1-Diff} ($1.20 \pm 0.01$ and $0.57 \pm 0.01$ respectively).
For any fixed
values of $\beta$ and $Q^{2}$, the pomeron and reggeon structure
functions were treated as free parameters to be fixed. Once
those parameters were determined for each set, it was possible to
calculate the ratio
\begin{equation}
R(\xi,\beta,Q^{2}) = {\int_{|t_{min}|}^{|t_{0, Lead}|}
F_{2}^{D(3)}(t,\xi,\beta,Q^{2})\,dt
\over \int_{|t_{min}|}^{|t_{0, Diff}|}
F_{2}^{D(3)}(t',\xi,\beta,Q^{2})\,dt'},
\end{equation}
which should be used to correct each measured diffractive structure function
data point at a given $\xi$, $\beta$ and $Q^{2}$.

Such procedure provided a correction factor that is a function of
$\xi$, going from 0.25 to 0.4. This is reflected in the curves shown
in Fig.~\ref{fig5}. From that figure it is clear
that the $f_{2}$ reggeon contribution coming from Fit~3 overestimates
the leading proton data by a factor 2 at least. The only parameter
related to this exchange is the normalization $N_{\tt I\!R}$, and
from Table~\ref{TableFit} it is clear that the fit to the diffractive
data alone drives such parameter to a very high value, compared with the
one from the global Fits 2 and 5, that are both quite compatible with
the combined sets of data.

\section{\label{Conclusions}Conclusions}

The analysis in this paper shows that we have to be very careful
before drawing conclusions about the role of Regge exchanges in
diffractive DIS. If only the H1 high statistic diffractive data
were used, as we have done in our Fit~3, an extrapolation of such a result
to the leading proton region will overestimate those data
by, at least, a factor 2 (Fig.~\ref{fig5}). It could be argued that such an
extrapolation
goes to low $\beta$ values beyond the range of the fitted data, and
our pomeron structure function would not be valid anymore. That is
true, but the point is that the pomeron contribution alone is not important
in such extrapolation. It is the secondary reggeon plus the pion
contribution that play the major role in the leading proton region.
The pion contribution itself is fixed and provides a quite reasonable
description of the leading neutron data. The same pion structure function is
used by the reggeon exchange, and it has been shown that a such
combination provides a good description of the leading baryon
data \cite{H1-Lead}. Therefore our choice of structure functions
for the secondary exchanges works well in both regimes, and it is fair
to expect that, extrapolating the information about the ratio
between pomeron and reggeon from the diffractive SF to the leading
proton regime, we should be able to have a decent qualitative description
of the leading proton data, but instead we were left with a result
that not only does not describe the data, but also lives no room for
corrections with extra reggeon exchanges.

The main problem in connecting the diffractive
and leading baryon regimes seems to come from the relative weight
that the fit put over the reggeon contribution in each case.
For instance, the normalization parameter $N_{\tt I\!R}$ changes from
$7.25$, when only diffractive data are used, to $2.058$, when both
diffractive and leading proton data are put together.

Although the interference term has some impact over
the reggeon contribution, it plays a minor role that does not improve
at all the discrepancies discussed above.

The fact that ZEUS Collab. has found no secondary exchange in their
diffractive measurements \cite{ZEUS-old,Zeus} is also an evidence that
the diffractive structure function data alone cannot conclusively
provide information concerning the contribution of the secondary
reggeon exchange in semi-inclusive $ep$ reactions.
Therefore, the leading baryon data represent an important constraint
that must be taken into account in any analysis based on the Regge
picture of diffraction.

The next step following this analysis is to show how these different
parameterizations affect the theoretical predictions for the cross
sections of diffractive photo- and electro-production of dijets, also
measured by ZEUS and H1 Collaborations \cite{Zeus-phot,H1-phot}. This
is going to be reported in a forthcoming paper \cite{Altem}.

\newpage

\section*{Acknowledgments} 
 
We would like to thank the Brazilian governmental agencies CNPq  
and FAPESP for financial support.


\begin{thebibliography}{99}

\bibitem{H1-old} H1 Collaboration, T. Ahmed {\textit et al.}, Phys. Lett. B  
{\textbf 348}, 681 (1995). 
 
\bibitem{ZEUS-old} ZEUS Collaboration, M. Derrick 
{\textit et al.}, Zeit. Phys. C {\textbf 68}, 569 (1995).  
 
\bibitem{H1-Lead}  H1 Collaboration, C. Adloff {\textit et al.}, Eur. Phys. 
J. C {\textbf 6}, 587 (1999). 
 
\bibitem{ZEUS-Lead} Nicol\`o Cartiglia, \eprint{hep-ph/9706416}. See also the 
information on physical results in section {\textit diffraction} of the 
ZEUS web page, \url{http://www-zeus.desy.de/publications.php3}.
 
\bibitem{Collins} P. D. B. Collins, {\textit An Introduction to Regge Theory  
and High Energy Physics}, (Cambridge University Press, Cambridge,  
England, 1977). 
 
\bibitem{Mirian}  M. Batista and R. J. M. Covolan, Phys. Rev. D {\textbf 59},  
054006 (1999). 

\bibitem{Szczurek} A. Szczurek, N. N. Nikolaev and J. Speth, 
Phys. Lett. B {\textbf 428}, 383 (1998). 

\bibitem{H1-Diff}  H1 Collaboration, C. Adloff {\textit et al.}, Z. Phys. C  
{\textbf 76}, 613 (1997). 
 
\bibitem{Zeus}  ZEUS Collaboration, J. Breitweg {\textit et al.}, Eur. Phys. 
J. C {\textbf 1}, 81 (1998). 
 
\bibitem{Ingelman}  G. Ingelman and P. Schlein, Phys. Lett. B 
{\textbf 152}, 256 (1985). 
 
\bibitem{Varios} R. J. M. Covolan and M. S. Soares, Phys. Rev. D 
{\textbf 57},  
180 (1998); L. Alvero {\textit et al.}, Phys. Rev. D 
{\textbf 59}, 074022 (1999);  
M. F. McDermott and G. Briskin, Proceedings of the  
Workshop ``Future Physics at HERA'', eds. G. Ingelman, A. De Roeck and R.  
Klanner, DESY, Hamburg, 1996, hep-ph/9610245; R. J. M. Covolan and
M. S. Soares, Phys. Rev. D  {\textbf 60}, 054005  (1999);
R. J. M. Covolan and M. S. Soares, Phys. Rev. D {\textbf 61}, 019901(E) (2000).  

\bibitem{Golec-Biernat} K. Golec-Biernat, J. Kwieci\'nski and
A. Szczurek, Phys. Rev. D {\textbf 56}, 3955 (1997).

\bibitem{F2param} H1 Collaboration, T. Ahmed {\textit et al.},
Nucl. Phys. {\textbf B439}, 471 (1995); H1 Collab, S. Aid {\textit et al.},
Nucl. Phys. {\textbf B470}, 3 (1996).

\bibitem{QCDNUM} {\textit QCDNUM16: A fast QCD evolution program},
M. A. J. Botje; Zeus Note 97-066. We are using QCDNUM version 16.10-12
in this work.

\bibitem{Royon} C. Royon, L. Schoeffel, J. Bartels, H. Jung,
R. Peschanski, Phys. Rev. D {\textbf 63}, 074004 (2001)

\bibitem{Dino1}  R. J. M. Covolan, J. Montanha and K. Goulianos, Phys. Lett.
B {\textbf 389}, 176 (1996).

\bibitem{Dino2}  K. Goulianos and J. Montanha, Phys. Rev. D 
{\textbf 59}, 114017 (1999).

\bibitem{GRV}  M. Gl\"{u}ck, E. Reya, A. Vogt, Z. Phys. C 
{\textbf 53}, 651 (1992).

\bibitem{Altem} R. J. M. Covolan, J. Montanha, A. N. Pontes and
M. S. Soares, in preparation.

\bibitem{Zeus-phot} ZEUS Collaboration, J. Breitweg {\textit at al.},
Eur. Phys J. C {\textbf 5} 41 (1998).

\bibitem{H1-phot} H1 Collaboration, C. Adloff {\textit et al.},
Eur. Phys. J. C {\textbf 6} 421 (1999); Eur. Phys. J. C {\textbf 20} 29 (2001).

\end{thebibliography}
\end{document}